\documentclass[aps,floats,epsf,twocolumn]{revtex4}
 \usepackage{graphics}

\begin{document}

\title{Pseudo-magnetic catalysis of the time-reversal-symmetry breaking in graphene}

\author{Igor F. Herbut}

\affiliation{Department of Physics, Simon Fraser University,
 Burnaby, British Columbia, Canada V5A 1S6}

\begin{abstract} A finite flux of the (time-reversal-symmetric) pseudo-magnetic field, which would represent the effect of a bulge in the graphene sheet for example, is shown to be a catalyst for spontaneous breaking of the time-reversal symmetry of Dirac fermions in two dimensions. Possible experimental consequences of this effect for graphene are discussed.
\end{abstract}
\maketitle

\vspace{10pt}

\section{Introduction}

It is well appreciated that the graphene sheet provides a particularly simple and physically relevant table-top  realization of the two-dimensional (pseudo) relativistic electron dynamics \cite{rmp}. In particular, the Dirac nature of graphene's quasiparticles provides these low-energy excitations with an extra protection from the usual effects of electron-electron interactions. The semi-metallic, non-interacting ground state of electrons in graphene may be understood as a Gaussian fixed point in the space of coupling constants, stable in all directions. Nevertheless, the ground state can in principle be turned into one with a broken symmetry at a finite, and, relative to the bandwidth, typically large interaction \cite{herbut1}. This way, for example, the system would acquire a finite staggered density, or a staggered magnetization, at a large nearest-neighbor and on-site repulsion, respectively. Both of these order parameters correspond to finite "masses" of the Dirac fermions that reduce the chiral ("valley", or "pseudo-spin") $SU(2)$ symmetry of the linearized Hamiltonian down to $U(1)$. The two-dimensional Dirac Hamiltonian, however, admits an additional mass-term that is invariant under the chiral symmetry, but odd under time-reversal \cite{haldane}. It has been argued recently that such a time-reversal-symmetry-breaking mass would be generated dynamically at a large second-nearest-neighbor repulsion between electrons on a honeycomb lattice \cite{raghu}. The type of mass, or an order parameter, that would eventually open up at strong coupling, seems to depend therefore on the non-universal details of the interactions on the atomic scale.

   Increasing the density of low-energy states is expected to enhance the effects of interactions on Dirac fermions. A manifestation of this general principle is the "magnetic catalysis", by which the chiral-symmetry-breaking mass is induced at an infinitesimal favorable interaction in a uniform magnetic field \cite{catalysis}, \cite{khveshchenko}. This mechanism is at the heart of several recent theories of some of the quantum Hall effects observed in graphene \cite{gusynin, herbut2, fuchs}. Magnetic field {\it cannot} catalyze the time-reversal-symmetry-breaking mass, however \cite{equation}. The purpose of this work is to show that the flux of the (non-abelian) pseudo-magnetic field plays the role of such a catalyzer. I demonstrate that in the presence of a finite flux of the non-abelian gauge field an infinitesimal favorable interaction would lead to the spontaneous breaking of the time-reversal symmetry of the ground state of two-dimensional Dirac fermions. This result is a general property of Dirac fermions in two dimensions, and as such it is independent of the specific nature of the underlying physical system. Nevertheless, its immediate significance derives from the notion that a component of such a pseudo-magnetic field represents the effect of smooth height variations of graphene's surface on the electron dynamics \cite{rmp}. With this possible application in mind I consider both the limits of a perfectly uniform and a spatially localized pseudo-magnetic flux. It is found that even the latter catalyzes a finite, but local, time-reversal-symmetry-breaking mass. Experimental conditions for  this non-intuitive manifestation of the coupling between the electronic and mechanical degrees of freedom in graphene are discussed.
   
\section{Dirac Hamiltonian and the time-reversal operator for graphene}

Let me establish the notation first.
   Consider the Dirac Hamiltonian for the four-component massless fermions in two spatial dimensions:
   \begin{equation}
   H[A^0, A ] = i \gamma_0 \gamma_i (p_i - A_i ^0 - A_i),
   \end{equation}
   where the repeated index $i=1,2$ is summed over, and $A_i$ is the general non-abelian $SU(2)$ gauge field
   \begin{equation}
   A_i = A_i ^3 \gamma_3 + A_i ^5 \gamma_5 + A_i ^{35} \gamma_{35}
   \end{equation}
   where $\gamma_{35} = i \gamma_3 \gamma_5$. $A_i ^0$ is the $U(1)$ (abelian) component that represents the physical magnetic field, whereas $A_i ^j$, $j=3,5,35$ multiply the three generators of the chiral $SU(2)$ symmetry \cite{dwave}
   of the free Dirac Hamiltonian $H[0,0]$. The five gamma-matrices satisfy $\{\gamma_\mu, \gamma_\nu \} = 2 \delta_{\mu \nu}$, $\mu=0,1,2,3,5$, and we will define them here to be all Hermitian. In our units, $\hbar=e=c=1$.

      The general mass-term that can be added to the Hamiltonian (1) which violates the $SU(2)$ chiral symmetry is given by $M=\vec{m}\cdot \vec{M}$, where $\vec{M}=(\gamma_0, i\gamma_0\gamma_3, i\gamma_0\gamma_5)$ is a vector under the chiral transformations. An additional mass-term  may then be defined to be a chiral scalar: $\tilde{m} \tilde{M}$, with $\tilde{M}= i\gamma_1 \gamma_2$. It is easy to check that the set of  all linearly independent matrices that anticommute with the free Dirac Hamiltonian $H[0,0]$ is exhausted by $\vec{M}$ and $\tilde{M}$, which therefore represent all the possible mass-terms.

      An important role in the discussion will be played by the time-reversal symmetry of the free Dirac Hamiltonian.
      As usual, the time-reversal is represented by an anti-unitary operator $I_t = U_t K$, where $U_t$ is unitary, and $K$ stands for the complex conjugation \cite{wigner}. Although everything that will be discussed hereafter will be  manifestly representation-independent, to exhibit the time-reversal operator one needs some representation of the $\gamma$-matrices. We prefer the "graphene representation" introduced earlier \cite{herbut1}, in which  $\gamma_0= I_2 \otimes \sigma_z$, $\gamma_1 = \sigma_z \otimes \sigma_y$, $\gamma_2 = I_2 \otimes \sigma_x$, $\gamma_3= \sigma_x \otimes \sigma_y$, and $\gamma_5 = \sigma_y \otimes \sigma_y$, with $\{ I_2, \vec{\sigma} \}$ as the standard Pauli basis in the space of two-dimensional matrices.  In this representation the time-reversal invariance of the free Dirac Hamiltonian $H[0,0]$ and of the general chiral-symmetry-breaking mass $M$ determines the unitary part of the time-reversal operator {\it uniquely} to be:
\begin{equation}
U_t= i\gamma_1 \gamma_5 = (\sigma_x \otimes I_2).
\end{equation}
 Postulating time-reversal invariance of both $H[0,0]$ and $M$ is motivated by the fact that these operators represent the low-energy limit of a completely real lattice Hamiltonian \cite{semenoff}.  As an immediate consequence, the chiral-symmetry preserving mass $\tilde{M}$ must be {\it odd} under time-reversal. This then is also in accord with the concrete lattice realization \cite{haldane} of  $\tilde{M}$.
 
 \section{Magnetic catalysis of chiral symmetry breaking}

   We begin by reformulating the mechanism of the magnetic catalysis in purely algebraic terms. Consider the Hamiltonian $H[A^0,0]$ with $A^0 \neq 0$. By virtue of representing the physical magnetic field $H[A^0,0]$ has the time-reversal symmetry broken, but the chiral symmetry preserved. In general, the spectrum of $H[A^0,0]$ will contain states with exactly zero energy \cite{aharonov}. Let us denote that zero-energy subspace of the full Hilbert space ${\cal H}_0$.  ${\cal H}_0$ is invariant under the generators of the chiral symmetry which by definition all commute with $H[A^0,0]$, but also under the operators that anticommute with $H[A^0,0]$, such as $\vec{M}$ and $\tilde{M}$. If we denote the trace of an operator {\it within} ${\cal H}_0$ as $Tr_0$, it follows that
   \begin{equation}
  Tr_0 \vec{M} = 0.
  \end{equation}
   This is because for each component of $\vec{M}$ there exists an operator which leaves ${\cal H}_0$ invariant and anticommutes with it \cite{remark}. In the basis of ${\cal H}_0$ which diagonalizes a chosen component of $\vec{M}$ the number of states with the eigenvalue $+1$ is thus equal to the number of those with the eigenvalue $-1$. Since one can write the ground state expectation value of a traceless operator that anticommutes with the Hamiltonian, such as $\vec{M}$, as \cite{semenoff1}, \cite{herbut3},
  \begin{equation}
  \langle  \vec{M} \rangle = \frac{1}{2} [ \sum_{n, occup} -
   \sum_{n, empty}]  \Phi_{0,n}^\dagger (\vec{x}) \vec{M} \Phi_{0,n} (\vec{x}),
  \end{equation}
  with $\{\Phi_{0,n}\}$ as a basis in ${\cal H}_0$, we see that occupying all the $+1$ zero-energy eigenstates and leaving the $-1$ eigenstates empty creates the maximal spatial average of the above order parameter.  At half-filling and in the non-interacting system, of course, the ground state is highly degenerate, and averaging over all the ground states ultimately leads to vanishing order. Nevertheless, in presence of even an infinitesimal interaction that favors a finite $\langle \vec{M} \rangle $, the non-interacting ground state is unstable towards a new non-degenerate ground state with all $+1$ states shifted slightly downward and all $-1$ states upward in energy, so that the chiral symmetry would become spontaneously broken.

      In a uniform magnetic field the above mechanism leads to a constant chiral-symmetry-breaking order parameter, and a gap in the spectrum at an infinitesimal favorable interaction between Dirac fermions, i. e. "magnetic catalysis" \cite{catalysis}, \cite{semenoff1}. Obviously the mechanism is quite general, and as will be discussed here it will be operative even if the magnetic field is not uniform, as long there is a finite support of the energy spectrum at zero.

  Before turning to our main subject, it is instructive to see why the above mechanism does not lead to the catalysis of the chirally symmetric order parameter  $\langle\tilde{M}\rangle $. First, note that unlike $\vec{M}$, $\tilde{M}$ commutes with {\it all} the other operators that leave ${\cal H}_0$ invariant, i. e. the generators of $SU(2)$ and $\vec{M}$, so it does not readily follow that its trace within ${\cal H}_0$ must vanish. In fact, since
  \begin{equation}
  H^2 [A^0, 0]= (p_i - A_i ^0)^2 + \tilde{M} \epsilon_{ij}\partial_i A_j^0,
  \end{equation}
  at least for an uniform (and say, positive) magnetic field  it is obvious that all states in ${\cal H}_0$ have the same ($-1$) eigenvalue of $\tilde{M}$. That this is generally true may be seen by rewriting the Dirac Hamiltonian in the magnetic field and in the Coulomb gauge $\partial_i A_i ^0 =0$ as
  \begin{equation}
  H[A^0,0] = e^{-\chi(\vec{x}) \tilde{M}}  H[0,0]  e^{-\chi(\vec{x}) \tilde{M}},
  \end{equation}
  where $A_i ^0 = \epsilon_{ij} \partial_j \chi$. This (non-unitary) transformation tells us that the zero-energy states of $H[A^0 , 0]$ and of the free Hamiltonian are related as
  \begin{equation}
  \Phi_{0,n} [A^0] (\vec{x}) \propto e^{\chi(\vec{x}) \tilde{M}} \Phi_{0,n} [0] (\vec{x}).
  \end{equation}
  Since for a total flux $F$ (in units of $hc/e$) localized near the origin, at large $|\vec{x}|$ $\chi(\vec{x})= F \ln|\vec{x}|$, the last equation implies that only the zero-energy eigenstates of $H[0,0]$ with the eigenvalue $-1$ of $\tilde{M}$ may lead to normalizable states of $H[A^0,0]$. All the states in ${\cal H}_0$ are thus the $-1$ eigenstates of $\tilde{M}$, even for an {\it arbitrary} configuration of the magnetic field.  Eq. (5) then implies  that $\int d \vec{x} \langle  \tilde{M} \rangle =0$ at half-filling for {\it any} occupation of the zero-energy states.

     To summarize, at the  filling one-half, the ground state of the Dirac Hamiltonian $H[A^0,0]$ in Eq. (1) in presence of a finite magnetic flux, which breaks the time-reversal and preserves the chiral symmetry, is inherently unstable towards the dynamical generation of the mass, that would break the chiral while preserving the time-reversal symmetry. I show next that when the physical (abelian) magnetic field vanishes and only the (non-abelian) pseudo-magnetic field is present, the same may be said, only with the "time-reversal" and the "chiral symmetry" in the last sentence exchanged.
     
 \section{Pseudo-magnetic catalysis of time-reversal symmetry breaking}

       The time-reversal symmetry, being broken by the magnetic field, did not play any role in the above discussion of the usual magnetic catalysis. Let us consider now the Hamiltonian $H[0,A]$ in Eq. (1), with $A \neq 0$. Since the time-reversal operator $I_t$ anticommutes with all the generators of $SU(2)$, it follows that $H[0,A ]$ is {\it even} under time-reversal.
   For a general non-abelian gauge configuration $A$ the chiral $SU(2)$ symmetry of the free Hamiltonian will be completely broken, and if $A$ is everywhere proportional to one and the same linear combination of the generators, it will be reduced to $U(1)$.  $\tilde{M}$, however, still always anticommutes with $H[0,A]$. ${\cal H}_0$ in this case will thus still be invariant under  $\tilde{M}$, as well as under $I_t$. As these two operators  anticommute, it immediately follows that when $A_i ^0 =0$ and $A_i \neq 0$,
   \begin{equation}
   Tr_0 \tilde{M}=0.
   \end{equation}
     Substituting $\tilde {M}$ for $\vec{M}$ in Eq. (5), it is now the chiral-symmetry-preserving, time-reversal-symmetry-breaking order parameter $\langle \tilde{M} \rangle$ that will become catalyzed in presence of an infinitesimal favorable interaction.

   Chiral-symmetry-breaking mass, in turn, is not catalyzed by the non-abelian gauge field. Assume for example  that $A_i = A_i ^{35} \gamma_{35}$, and $A_i ^0 =0$. Similar to Eq. (7) one can then write \cite{jackiw}
   \begin{equation}
  H[0,A] = e^{-\phi(\vec{x}) \gamma_0}  H[0,0]  e^{-\phi(\vec{x}) \gamma_0},
  \end{equation}
  where now $A_i ^{35} = \epsilon_{ij} \partial_j \phi$. In analogy with the Eq. (8) it follows that all the states in ${\cal H}_0$ now have the same eigenvalue of $\gamma_0$, and it is the chiral-symmetry-breaking order parameter $\int d\vec{x} \langle \gamma_0 \rangle$ that vanishes at half-filling.

     To see the dynamical consequences of the above algebra more explicitly, consider the Lagrangian density
     \begin{equation}
     {\cal L}=  \bar{\Psi}(x) \gamma_\mu (\partial_\mu - A_\mu ^{35} \gamma_{35} ) \Psi(x)  - \frac{g}{2} ( \Psi^\dagger (x) \tilde{M} \Psi(x))^2
     \end{equation}
     with an interaction $g>0$, $\mu=0,1,2$, $x=(x_0,\vec{x})$, $x_0$ as the imaginary time, and the quenched component of the non-abelian gauge field $A_i ^{35} (\vec{x}) \neq 0$. Introducing the Hubbard-Stratonovich field this can be rewritten as
     \begin{eqnarray}
     {\cal L}=  \bar{\Psi}(x) \gamma_\mu (\partial_\mu - A_\mu ^{35} (\vec{x}) \gamma_{35} ) \Psi(x)  + \\ \nonumber
      \frac{1}{2g}\tilde{m}^2 (x) - \tilde{m}(x) \Psi^\dagger (x) \tilde{M} \Psi (x).
     \end{eqnarray}
     The mean-field theory of the above interacting problem would amount to minimization of the corresponding action $\int {\cal L} dx$ with respect to $\tilde{m}(x)$, or equivalently, to determining the ground-state expectation value
     \begin{equation}
     \langle \Psi^\dagger (x) \tilde{M} \Psi (x)\rangle =  \langle \tilde{m} (x) \rangle/g,
     \end{equation}
     self-consistently. An uniform $\langle \tilde{m} (x) \rangle$ may be understood as the time-reversal symmetry breaking order parameter of ref. \cite{raghu}.
     For a constant pseudo-magnetic field $B^{35}= \partial _1 A_2 ^{35}- \partial _2 A_1 ^{35}$, in full analogy with the standard magnetic catalysis \cite{catalysis}, \cite{herbut2} we then find
\begin{equation}
\langle \Psi^\dagger (x) \tilde{M} \Psi (x)\rangle = B^{35} + O(g),
\end{equation}
where the first term derives from the split zero-energy level, and the term $O(g)$ is due the other Landau levels. For an inhomogeneous  $B^{35}(\vec{x})$ the self-consistent calculation can be performed only numerically. Here we circumvent this hurdle by dropping the self-consistency requirement and minimizing the action with respect to only a {\it uniform } $\tilde{m}$. This may be understood as a variational calculation, or as the exact solution of the Berlin-Kac version of the theory \cite{kac}, in which the contact interaction in Eq. (11) is replaced with the interaction of an infinite range  \cite{tesanovic}:
\begin{equation}
-\frac{g}{2\Omega } \int dy (\Psi^\dagger (x) \tilde{M} \Psi (x)) (\Psi^\dagger (y) \tilde{M} \Psi (y) ),
\end{equation}
 with $\Omega$ as the area of the system. The uniform ansatz becomes an exact solution of the modified theory in the thermodynamic limit $\Omega \rightarrow \infty$. In either case there is a gap of $2\tilde{m}$ in the spectrum, which satisfies
\begin{equation}
\frac{\tilde{m}}{g} = \frac{F}{\Omega} + \tilde{m} \int_0 ^\infty \frac{{\cal N}(\epsilon) d\epsilon}{ (\epsilon ^2 + \tilde{m} ^2)^{1/2} },
\end{equation}
 with ${\cal N} (\epsilon )$ as the exact density of states per unit area of the non-interacting Dirac fermions in the flux of $A_i ^{35}$, at $\epsilon \neq 0$. Since at low energies we expect that  ${\cal N}(\epsilon) \propto \epsilon ^{(2-z)/z}$  \cite{ludwig}, for $z< 2$ the second term may be neglected at a weak coupling, and $\tilde{m}$ is finite in the thermodynamic limit only in the case of an extensive flux, $F\propto \Omega$ \cite{remark1}. Nevertheless, even if $F$ is finite the expectation value of the time-reversal symmetry breaking order parameter is finite and equal to
 \begin{equation}
 \lim_{\Omega\rightarrow \infty} \langle \Psi^\dagger (x) \tilde{M} \Psi (x)\rangle= \frac{1}{2}
 \sum_{{\cal H}_0} \Phi^\dagger _{0,n} (\vec{x})  \Phi _{0,n} (\vec{x}).
 \end{equation}
  A finite pseudo-flux selects the time-reversal-symmetry broken ground state out of the degenerate manifold, in close parallel with the standard formalism of spontaneous symmetry breaking in statistical physics \cite{book}.

 To illustrate the local character of the order parameter for finite $F$, let us exhibit the sum in the last equation  for the particular pseudo-magnetic field
   \begin{equation}
   B^{35}(r)= \frac{2F}{ R^2 (1+(r/R)^2)^2 }.
   \end{equation}
 In the graphene representation the zero-energy state with $\pm 1$ eigenvalue of $\tilde{M}$ are then
   $\Phi_{n,-} ^\dagger (\vec{x}) = f^* _n (\vec{x}) (0,1,0,0)$, and $\Phi_{n,+} ^\dagger  (\vec{x}) = f_n (\vec{x}) (0,0,0,1)$, where
   \begin{equation}
   f_n (\vec{x})= \frac{ \pi^{-1} R^{-2(n+1)}  (x+iy)^n}{ \sqrt{ \beta(n+1,F-n-1) (1+ (r/R)^2 ) ^F } },
   \end{equation}
   with the integer $n <F$. Note that the $\Phi_{n,+}= I_t \Phi_{n,-}$.
   For an integer flux $F$ then the sum in Eq. (17) can be exactly performed with the result
   \begin{equation}
 \lim_{\Omega \rightarrow \infty} \langle \Psi^\dagger (\vec{x}) \tilde{M} \Psi (\vec{x})\rangle=
   (1-F^{-1} ) B^{35} (r).
 \end{equation}
  For a general localized flux the precise proportionality between the order parameter and the field obtains only in the limit $F \gg 1$ \cite{dunne}. The order parameter, however, is always localized in the region of flux.
  
 \section{Experimental consequences} 

  Finally, let us address possible consequences of the above results for graphene. As mentioned in the introduction, the time-reversal-symmetry-breaking mass is favored by the second-nearest-neighbor repulsion \cite{raghu},
      whereas the competing chiral-symmetry-breaking masses are preferred  by the nearest-neighbor repulsion between electrons. With the electron spin included, chiral-symmetry-breaking mass with the opposite sign for the two spin components, which corresponds to  staggered magnetization, is also preferred by the, most likely the strongest, on-site repulsion \cite{herbut1}. As one has little control over the size of the interaction couplings and can hope only to alter the bandwidth, the possible instability towards the time-reversal-symmetry-breaking mass without any gauge fields seems likely to be inferior to the one towards chiral-symmetry breaking. An "application"  of the pseudo-magnetic flux, however, changes this, since it is {\it only} the time-reversal-symmetry-breaking mass that is catalyzed by it at weak interactions.

        A crude estimate of the locally catalyzed gap gives $\tilde{m}\approx V B^{35} / B_{latt}$, where $B_{latt}\approx 10^4 T$ is the characteristic lattice magnetic field scale, and $V$ is the strength of the second-nearest-neighbor repulsion. A (single) wrinkle which tends to spontaneously form on a graphene sheet would already lead to $B^{35} \sim 1T $ \cite{morozov}, \cite{vafek}, so together with an estimate of $V\sim (1 - 5) eV$ \cite{herbut1},  $\tilde{m}\sim (0.1-0.5) meV$. A randomly wrinkled graphene corresponds to zero total flux, of course, and so $\langle \tilde{m}(\vec{x}) \rangle =0$. To produce a finite net pseudo-magnetic flux one needs to deliberately bulge the graphene sheet, which according to the above estimate should push the gap well into the meV range.

 The pseudo-magnetic catalysis described here is stable with respect to deviations from half-filling, i. e. for the chemical potential smaller than the generated mass.
 
 \section{Summary}

 To conclude, I described the mechanism complementary to the usual magnetic catalysis: a finite net flux of a component of the non-abelian gauge field, which preserves the time-reversal and breaks the chiral symmetry of the free Dirac Hamiltonian, serves as a catalyst of the time-reversal-symmetry-breaking, chiral-symmetry-preserving order parameter. This could lead to local spontaneous breaking of the time reversal symmetry in graphene where such a pseudo-magnetic field is provided by a bulge in graphene's plane, due to the second-nearest-neighbor repulsion term in the lattice Hamiltonian. The magnitude of the effect should be large enough for the gap in the local density of states to become observable by scanning tunneling microscopy, for example \cite{li}.
 
 \section{Acknowledgement} 

   This work is supported by the NSERC of Canada.


\begin{thebibliography}{99}
\bibitem{remark} For the same reason $Tr_0 \Gamma =0$, where $\Gamma$ is a generator of the chiral $SU(2)$.
\end{thebibliography}
\end{document}